\documentclass[aps,prb,notitlepage,showkeys,showpacs,superscriptaddress,nofootinbib,longbibliography,preprintnumbers]{revtex4-2}

\usepackage[utf8]{inputenc}

\usepackage{tikz}
\usetikzlibrary{intersections,calc,arrows.meta}

\usepackage{here}
\usepackage{ulem}%取り消し線

\usepackage{braket}

\usepackage{amscd}%可換図式

\usepackage[top=30truemm,bottom=30truemm,left=25truemm,right=25truemm]{geometry}

\usepackage{spectralsequences}
\usepackage{tikz}

\usepackage{xcolor}
\newcommand{\Z}{\mathbb{Z}}
\newcommand{\tr}[1]{\mathrm{tr}\left(#1\right)}
\newcommand{\coho}[3]{\mathrm{H}^{#1}(#2;#3)}
\newcommand{\cohoZ}[2]{\mathrm{H}^{#1}(#2;\mathbb{Z})}
\usepackage[all]{xy}

\usepackage[colorlinks=true,linktoc=page,citecolor=red,linkcolor=blue]{hyperref}	
\usepackage[whole]{bxcjkjatype} % for Japanese

\usepackage{slashed}

\usepackage{amsthm}
\usepackage{thmtools}
\usepackage{blindtext}
\usepackage{braket}

\declaretheoremstyle[
       shaded={bgcolor=\color{rgb}{0.9,0.9,0.9}}  % comment this line in/out
]{theorem}

\declaretheoremstyle[
       shaded={bgcolor=\color{rgb}{0.9,0.9,0.9}}% comment this line in/out
]{question}

\declaretheoremstyle[
       shaded={bgcolor=\color{rgb}{0.9,0.9,0.9}}  % comment this line in/out
]{remark}

\declaretheoremstyle[
       shaded={bgcolor=\color{rgb}{0.9,0.9,0.9}}  % comment this line in/out
]{proposition}

\declaretheoremstyle[
       shaded={bgcolor=\color{rgb}{0.9,0.9,0.9}}  % comment this line in/out
]{definition}

\declaretheoremstyle[
       shaded={bgcolor=\color{rgb}{0.9,0.9,0.9}}  % comment this line in/out
]{assumption}

\declaretheoremstyle[
       shaded={bgcolor=\color{rgb}{0.9,0.9,0.9}}  % comment this line in/out
]{conjecture}

\declaretheoremstyle[
       shaded={bgcolor=\color{rgb}{0.9,0.9,0.9}}  % comment this line in/out
]{corrorary}

\declaretheoremstyle[
       shaded={bgcolor=\color{rgb}{0.9,0.9,0.9}}  % comment this line in/out
]{axiom}

\declaretheoremstyle[
       shaded={bgcolor=\color{rgb}{0.9,0.9,0.9}}  % comment this line in/out
]{lemma}

\usepackage{mathrsfs}
\usepackage{amsmath,amssymb}

\usepackage{esvect}

\begin{document}

\title{Higher structures in matrix product states}

\author{Shuhei Ohyama}
\email{shuhei.oyama@yukawa.kyoto-u.ac.jp}
 \affiliation{Center for Gravitational Physics and Quantum Information, Yukawa Institute for Theoretical Physics, Kyoto University, Kyoto 606-8502, Japan}

\author{Shinsei Ryu}
%\thanks{The authors are listed in alphabetical order.}
\affiliation{Department of Physics, Princeton University, Princeton, New Jersey, 08544, USA}

\date{\today} 
\preprint{YITP-23-39}

\begin{abstract}
For a parameterized family of invertible states 
(short-range-entangled states) in $(1+1)$ dimensions, 
we discuss a generalization of the Berry phase.
Using translationally-invariant, 
infinite matrix product states (MPSs), 
we introduce a gerbe structure, 
a higher generalization of complex line bundles, 
as an underlying mathematical 
structure describing topological properties of a parameterized family of matrix product states.   
We also introduce a "triple inner product" 
for three matrix product states,
which allows us to extract a topological invariant, 
the Dixmier-Douady class over the parameter space. 
\end{abstract}

\maketitle

\tableofcontents

\section{Introduction}

\subsection{The Berry Phase and Its Higher Generalization}

Quantum mechanical phase degrees of freedom are known to have an interesting interplay with topology 
\cite{PhysRev.115.485,1931RSPSA.133...60D}.
A canonical example is the Dirac monopole 
where the presence of a magnetic monopole prevents quantum mechanical wave
functions from being defined uniquely over the entire space. Instead, wave
functions can be defined by introducing multiple patches, and at the intersection of two patches, wave functions from different patches are related
by a transition function \cite{1975PhRvD..12.3845W}. The (large) gauge
invariance results in the quantization of magnetic charges in units of the inverse of the fundamental charge.  A magnetic monopole also arises in the
context of the Berry phase, where a diabolic point of the Hamiltonian plays the
role of the Dirac monopole of the Berry connection in a parameterized quantum system 
where the wave function $|\psi(x)\rangle$ 
depends smoothly on some adiabatic parameter(s) $x$ taken from a parameter space $X$. 
The mathematical structure underlying these situations is a principle U(1) bundle over the parameter space $X$.
Such bundles are  characterized and classified by a topological invariant,
the first Chern class taking its value in the second cohomology group of $X$,
$\cohoZ{2}{X}$.

The Berry phase also plays an important role in topological phenomena in many-body quantum physics
such as quantum Hall states and Chern insulators \cite{TKNN82, Kohomoto85} and 
the Thouless pump
\cite{PhysRevB.27.6083}.
An important class of topological states
is the so-called invertible states 
(short-range-entangled states)
that are
realized as a unique ground state of a gapped Hamiltonian. 
Invertible states can be protected by symmetry from
being topologically trivial 
(symmetry-protected topological (SPT) phases), 
as known in topological insulators and the Haldane spin chain 
\cite{RevModPhys.83.1057, RevModPhys.82.3045, HALDANE1983464, PhysRevLett.50.1153}.
Symmetry-protected (and often discrete) Berry phases
are important in characterizing these phases\footnote{
The Berry phase or geometrical phase is commonly discussed as 
a phase that quantum wavefunction acquires during adiabatic time evolution. Unlike the overall phase of quantum mechanical wavefunctions, which is unobservable, the Berry phase has observable consequences.
In this paper, we broaden the usage of the term "Berry phase" to indicate  
the phases of wavefunction overlaps that may encode 
topological information of topological states and processes.
For instance, in SPT phases, the discrete phases acquired by wavefunctions through non-adiabatic discrete transformations are often discussed as topological invariants. 
(See, for example, discrete partial rotation used in Ref.\ \cite{Shiozaki_2017}.)
Also, in Wu-Yang's work on magnetic monopoles,
the transition functions connecting wavefunctions 
from different patches are physical and determine 
the topological class (the first Chern class). 
In this paper, we loosely call 
these phases associated with wavefunction overlap
the Berry phase, although in this description of magnetic monopoles, we do not need the Berry connection. 
In a similar vein, by the higher Berry phase,
we mean 
the phase of the triple inner product of wavefunctions (defined below) without explicitly using a (higher generalization of) Berry connection.
It determines the topological class (the Dixmier-Douady class)
of a family of invertible states over $X$.}.
There are however many-body systems where the regular notion of the Berry phase 
fails to capture topological properties.
In recent years, 
a family of invertible states 
that depends on some parameter $x \in X$
has been discussed
\cite{KS20-1,KS20-2,
Hsin_2020,
Cordova_2020a, Cordova_2020b,
shiozaki21,
Choi_2022,
OTS23,
beaudry2023homotopical}.
Such a family can be topologically non-trivial and can be considered 
as a generalization of the Thouless pump. 
It can also be considered as a generalization 
of regular gapped phases (SPT phases)
where the parameter space is a single point. 
For example, it is known that there is a nontrivial family of $(d+1)$-dimensional systems with $\mathrm{U}(1)$ symmetry parameterized 
over
$S^d$
\cite{KS20-2}. 
We however cannot use the ordinary Berry phase to detect its nontriviality in general. 
A cursory explanation is that
the Berry connection and Berry curvature measure the nontriviality of $\cohoZ{2}{X}$, so for example when $d=3$, they cannot be nontrivial on $S^3$.
Even worse, 
if not introduced carefully, 
the Berry connection and curvature 
may be ill-defined
in  
many-body quantum systems 
in the first place:
For example, 
if we 
consider a chain of spins
that are weakly interacting with each other 
and are each coupled to an adiabatically time-evolving magnetic field, the $1$st Chern number diverges in the thermodynamic limit since 
each spin contributes independently.
We could instead consider 
the Chern number per unit cell, 
but it is not necessarily quantized in general.

In order to capture the topology of 
higher generalizations of the Thouless pumping, 
it has been realized that a "higher" generalization of the Berry phase, which takes its value in $\cohoZ{d+2}{X}$, is important \cite{Kitaev13,KS20-2,OTS23}. 
The purpose of this paper is to extend the ordinary Berry phase
to $(1+1)$-dimensional quantum many-body systems motivated by these trends, and construct a topological invariant that takes its value in $\cohoZ{3}{X}$. 
In this paper, 
the families of invertible states we consider do not 
preserve some symmetry, 
e.g., particle number conserving U(1).

\subsection{Summary of The Paper}

In this paper, we identify a gerbe structure 
for parameterized families of invertible states 
in $(1+1)$ dimensions using translationally invariant, infinite matrix product states (MPSs).
A gerbe is a higher generalization of complex line bundles 
and provides, as we will see, 
a natural framework to discuss the higher Berry phase. 
(We will give a brief overview of a gerbe 
in Sec.\ \ref{What Is A Gerbe and Why?}.)
We will show how we can construct a gerbe from 
a family of infinite MPSs.
We also show how the data constituting the gerbe,
and its topological invariant in particular, 
can be extracted from a (properly generalized) 
overlap of three MPSs.
We call the overlap 
the triple inner product,
which is depicted in Fig.\  \ref{fig:boomerang}.
This is analogous to Wu-Yang's work 
where we can extract the ordinary Berry phase by taking the inner product 
of two wavefunctions that are physically 
the same but taken from two different patches. 
In our generalization, we extract the "higher" Berry phase by taking 
the
"triple inner product" of the three physically same states in three different patches. 
This "triple inner product" gives the Dixmier-Douady class over the parameter space $X$ that takes its value in $\cohoZ{3}{X}$. 
Our formalism works both for 
the torsion and free parts of 
$\cohoZ{3}{X}$, 
i.e., the cases when 
families of invertible states over $X$
are classified by
a finite order group
or (copies of) the cyclic group $\mathbb{Z}$, 
respectively.
For the free case, 
as we will discuss, 
we need to deal with 
MPSs whose rank (bond dimension)
is not constant over 
the parameter space $X$. 
Finally, we will also discuss this gerbe structure 
and the triple inner product are naturally described by using 
the language of non-commutative geometry, 
a star product and integration.

\section{Construction of a Gerbe from MPS}\label{sec:const_gerbe}

\subsection{Brief Review of MPS}\label{Brief Review of MPS}

\begin{figure}[t]
 \begin{center}
  \includegraphics[scale=1.5]{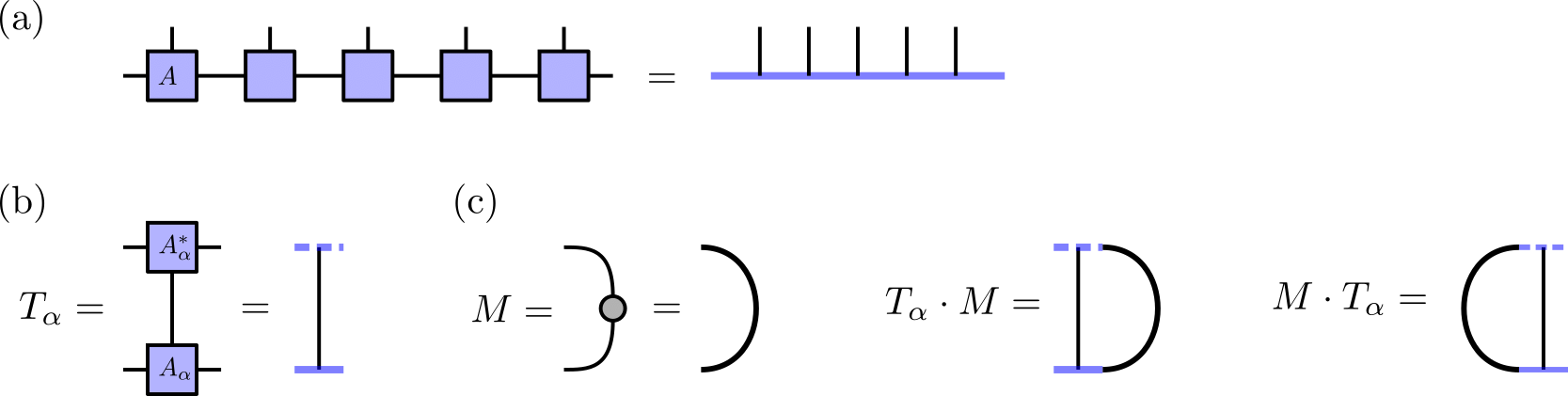}
 \end{center}
  \caption[]{
  Matrix product states (a),
  transfer matrices (b),
  and the left and right actions of 
  transfer matrices (c).
  When there is no confusion, 
  we simplify our notation by not 
  showing boxes representing tensors
  explicitly.
  The conjugate of the MPS matrix
  $A$ is represented by dotted lines.
}
\label{fig: action of transfer matrix}
\end{figure}

This paper focuses on invertible states 
(short-range entangled states) in $(1+1)$ dimensions. 
In particular, we will study 
families of translationally-invariant invertible states that depend on a parameter $x\in X$. 
Such a parameterized family can be called invertible states over $X$.
Invertible states in $(1+1)$ dimensions are efficiently represented 
as MPSs, so we begin by reviewing the necessary 
ingredients of MPSs.
Specifically, we will deal with translationally-invariant, infinite MPSs.
For a more in-depth discussion, 
see, for example, 
Refs.\ \cite{P-GVWC07, Vidal_2007,KZMBJP13,Cirac_2021}.

As a start, let us consider a finite one-dimensional 
lattice with $L$ sites,
labeled by $j=1,\cdots, L$.  
Let $\mathfrak{h}_{j}$ 
be a local Hilbert space with dimension $d$ 
(independent of $j$),
where $\{\ket{i}\}_{i=1}^{d}$ is an orthonormal basis 
of $\mathfrak{h}_{j}$. 
The total Hilbert space of the chain is 
$\mathcal{H}:=\bigotimes_{j=1}^L \mathfrak{h}_{j}$. 
A translationally invariant MPS is defined by 
a set of $n\times n$ matrices $\{A^{i}\}$ 
with the same index as the orthonormal basis. 
With periodic boundary conditions, 
the MPS generated by $\{A^{i}\}$ is given by
\begin{eqnarray}
    \ket{\{A^{i}\}}_{L}:=\sum_{\{i_{k}\}}\tr{A^{i_1}\cdots A^{i_L}}\ket{i_1, \ldots, i_L},
\end{eqnarray}
where $\sum_{\{i_{k}\}}$ represents a summation 
over all configurations of $(i_1,\ldots, i_L)$,
$\sum_{\{i_{k}\}}=\sum_{i_1}\cdots\sum_{i_L}$. 
MPSs with fixed boundary conditions can be defined similarly
with boundary vectors specifying boundary conditions.

We are interested in invertible states in the thermodynamic limit, $L\to \infty$,
where boundary conditions play no role.
In this limit, 
the physical properties of the MPS are encoded in its transfer matrix
which is defined by 
\begin{eqnarray}
T_A:=\sum_{i}A^{i\ast}\otimes A^{i}.
\end{eqnarray}
A transfer matrix $T_A$ acts on 
$M\in \mathrm{Mat}_{n}(\mathbb{C})$ from the left 
and right as
\begin{align}
  \label{eq:left_action}
T_A\cdot M:=\sum_{i}A^{i} M A^{i\dagger},
\\
\quad 
  \label{eq:right_action}
M\cdot T_A:=\sum_{i}A^{i\dagger}M A^{i},
\end{align}
respectively.
We represent these actions pictorially 
in Fig.\ \ref{fig: action of transfer matrix}. 

Invertible states are represented by an injective MPS,
which can be defined, using a transfer matrix, as follows
\cite{P-GVWC07}: 
Let $\{A^{i}\}$ be a set of $n\times n$ matrices 
and $r_{A}$ be the spectral radius 
of the transfer matrix. 
Then $\{A^i\}$ is injective if and only 
if the left action of the transfer matrix has a unique eigenvalue 
$\lambda$ with eigenvalue 
$\left|\lambda\right|=r_{A}$ 
and the eigenvector $\Lambda$ is 
a positive definite $n\times n$ matrix. 
We call an MPS generated by injective matrices 
an injective MPS. 
For injective matrices, 
it is known that the spectral radius $r'_A$ 
for the right action  
is equal to $r_A$, i.e., $r_A=r'_A$. In addition, 
a right eigenvalue $\lambda'$ with $\left|\lambda'\right|=r_A$ 
is unique and the corresponding eigenvector $\Lambda'$ 
is a positive definite matrix.

For injective matrices, 
the eigenvalue equation $T_A\cdot\Lambda=\lambda\Lambda$ 
can be rewritten as 
\begin{eqnarray}
    \sum_{i}A^{i}\Lambda A^{i\dagger}=\lambda\Lambda
    \quad
    \Longleftrightarrow
    \quad 
    \sum_i A^{i}_{\rm c}A^{i\dagger}_{\rm c}=1_n,
\end{eqnarray}
where $A^{i}_{\rm c}:=\frac{1}{\sqrt{\lambda}}\Lambda^{-\frac{1}{2}}A^{i}\Lambda^{\frac{1}{2}}$. 
We call $\{A^{i}_{\rm c}\}$ the right canonical form 
of the injective matrices $\{A^{i}\}$. 
In this form, the spectral radius for the left action 
\eqref{eq:left_action} is $1$, 
and
the eigenvector is 
modified, 
$\Lambda' \to \Lambda^{-\frac{1}{2}}\Lambda'\Lambda^{\frac{1}{2}}$, which is not the identity matrix in general.

In the following, unless otherwise mentioned, 
we take our MPSs
to be in the right canonical form and denote 
the eigenvectors with eigenvalue $1$
for the left and right actions as
$\Lambda^R_A$ and $\Lambda^L_A$, respectively:
\begin{align}
  T_A\cdot\Lambda^{R}_A=\Lambda^{R}_A,
                       \quad
  \Lambda^{L}_A \cdot T_A=\Lambda^{L}_A.
\end{align}
In the present case, $\Lambda^{R}_A$ is just the identity matrix, but in the later generalization, 
the case where it is not the identity matrix will appear, so we assign a symbol to it in advance.

\begin{figure}[t]
 \begin{center}
  \includegraphics[scale=1.5]{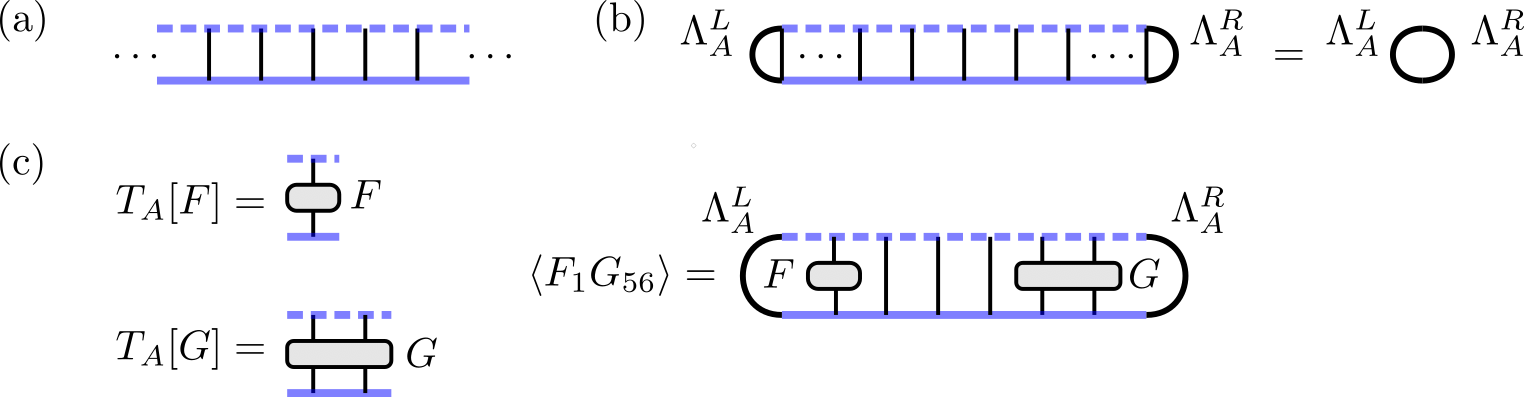}
 \end{center}
  \caption[]{(a) 
  The inner product (norm)
  of infinite MPSs 
  contains infinitely many products of the transfer matrices. 
  (b) In the thermodynamic limit, only the eigenvector with maximal eigenvalue survives. In the infinite MPS formalism, the inner product is defined by contracting the left end with the left eigenvector $\Lambda_{A}^{L}$ and the right end with the right eigenvector $\Lambda_{A}^{R}$. 
  By using the eigenvalue equation, this value is found to be equal to $\tr{\Lambda_{A}^{L}\Lambda_{A}^{R}}$. 
  (c) In general, the expectation value of a local observable is defined by putting the left eigenvector on the left side of the operator with the left-most support and the right eigenvector on the right side of the operator with the right-most support.
  }
  \label{fig:infiniteMPS}
\end{figure}

By using 
the
left and right eigenvectors $\Lambda_{A}^{L}$ and $\Lambda_{A}^{R}$, an infinite MPS is defined in the following manner
\cite{Vidal_2007,Vidal_2008,KZMBJP13}: For infinite systems, it is difficult to define the state itself, since 
an MPS on an infinite system is formally given by
\begin{eqnarray}
    \ket{\{A^{i}\}}_{\infty}:=\sum_{\{i_{k}\}}\cdots A^{i_1}\cdots A^{i_L}\cdots\ket{\cdots i_1\cdots i_L\cdots}
\end{eqnarray}
and its coefficients have an ambiguous infinite product of matrices. 
In the infinite MPS formulation, we give up defining the state itself but define the expectation value of the state. 
An expectation value of local observable contains infinitely many products of transfer matrices in the right and left directions (Fig.\ \ref{fig:infiniteMPS}). 
Therefore, in the infinite size limit, the product only has a value on the eigenvector space of the transfer matrix with the maximum eigenvalue. So, we close the right and left ends with $\Lambda_{A}^{L}$ and $\Lambda_{A}^{R}$ to define the expectation value. For example, the inner product of $\ket{\{A^i\}}_{\infty}$ is defined by
\begin{eqnarray}
    \braket{\{A^{i}\}|\{A^{i}\}}_\infty=\Lambda_{A}^{L}\cdot (T_{A})^{N}\cdot\Lambda_{A}^{R}=\tr{\Lambda_{A}^{L}\Lambda_{A}^{R}}.
\end{eqnarray}
for arbitrary $N\in\mathbb{N}$.
In the right canonical form, $\Lambda_{A}^{R}=1_n$ but the phase of $\Lambda_{A}^{L}$ is not fixed. As a normalization condition for the infinite MPS, we fix the phase of $\Lambda_{A}^{L}$ by $\tr{\Lambda_{A}^{L}}=1$.
Similarly, for example, the expectation value of local operators $F_1$ (acting on the site $1$) and $G_{56}$ (acting on the site $5$ and $6$) are given by
\begin{eqnarray}
    \braket{F_{1}G_{56}}:=\Lambda_{A}^{L}\cdot 
    T_A[F_1]
    (T_{A})^{3}
    T_A[G_{56}] \cdot \Lambda_{A}^{R},
\end{eqnarray}
where $(T_A[F_{1}])_{(a,c),(b,d)}:=\sum_{i,j} A^{i\ast}_{ab}F^{ij}_{1}A^{j}_{cd}$ 
and 
$(T_A[G_{56}])_{(a,d),(c,f)}:=\sum_{i,j,k,l}\sum_{b,e}A^{i\ast}_{ab}A^{j\ast}_{bc}G^{ij,kl}_{56}A^{k}_{de}A^{l}_{ef}$.

\subsection{What Is A Gerbe and Why?} 
\label{What Is A Gerbe and Why?} 

As we are interested in invertible states over $X$, 
we consider a family of infinite MPSs,
$\{A^i(x)\}$, where the corresponding transfer matrix, left and right eigenvectors, etc.
are also dependent on $x$.
We will call such a family as MPSs over $X$.

As mentioned in Introduction, 
a parameterized family of quantum mechanical states with ordinary Berry phase
can be described by a complex line bundle.  
Let us consider an open covering of $X$, $\{U_{\alpha}\}$.
A complex line bundle is defined by transition functions 
$e^{i2\pi\phi_{\alpha\beta}}$ on intersections 
$U_{\alpha\beta}:=U_{\alpha}\cap U_{\beta}$.
They satisfy $e^{i2\pi\phi_{\beta\alpha}} = e^{-i 2\pi \phi_{\alpha\beta}}$,
and also, on triple intersections $U_{\alpha\beta\gamma}:=U_{\alpha}\cap U_{\beta} \cap U_{\gamma}$,
\begin{eqnarray}\label{eq:cocycle_condi}
    e^{i2\pi\phi_{\alpha\beta}} e^{i2\pi\phi_{\beta\gamma} }e^{i2\pi\phi_{\gamma\alpha}} =1.
\end{eqnarray}
A transition function $e^{i2\pi\phi_{\alpha\beta}}$ is an element of 
the \v{C}eck complex $C^{1}(X;\underline{\mathrm{U}(1)})$ and Eq.\ (\ref{eq:cocycle_condi}) is nothing but the cocycle condition. 
Therefore, $e^{i2\pi\phi_{\alpha\beta}}$ defines a $1$st \v{C}eck cohomology class $\left[e^{i2\pi\phi_{\alpha\beta}}\right]\in\coho{1}{X}{\underline{\mathrm{U}(1)}}\simeq\cohoZ{2}{X}$\footnote{Here, the isomorphism is given in the following way: Let's take a $\mathbb{R}$-lift $\hat{\phi}_{\alpha\beta\gamma}$ of $\phi_{\alpha\beta\gamma}\in\mathbb{R}/\mathbb{Z}$. Then, on $U_{\alpha\beta\gamma}$,  $f_{\alpha\beta\gamma}:=\hat{\phi}_{\alpha\beta}-\hat{\phi}_{\alpha\gamma}+\hat{\phi}_{\beta\gamma}$ takes its value in $\mathbb{Z}$ and satisfies the cocycle condition. Thus it defines the $2$rd cohomology class $\left[f_{\alpha\beta\gamma}\right]\in\cohoZ{2}{X}$, and this is a topological invariant of the complex line bundle.},
and it is measured by the $1$st Chern class. 
Here, the underbar represents that it is the sheaf cohomology.

In this paper, we consider a higher generalization of the Berry phase
for a parameterized family of $(1+1)$-dimensional 
invertible states
\cite{KS20-1,Xueda21,OTS23}.
We expect that these higher generalizations of the Thouless pumping
can be topologically classified by
the Dixmier-Douady class that takes its value in $\cohoZ{3}{X}$.
We propose that the mathematical structure that describes
the topological classification of higher Thouless pumping is a gerbe.
A gerbe is a higher generalization of a complex line bundle.
It has been used to describe, for example,
the $(1+1)$-dimensional Wess-Zumino-Witten models, 
the $(2+1)$-dimensional Chern-Simons theories, 
the Kalb-Ramond $B$-field and D-branes in string theory, and
various anomalies in quantum field theory
\cite{Gawedzki_2005, Kapustin99, CAREY_2000}.
Let us first briefly introduce the mathematical definition of a gerbe.
In the next subsection, we will
then construct a gerbe from MPSs over $X$.

Let $X$ be a topological space. A gerbe on $X$ is described by datum $(\{U_{\alpha}\},\{L_{\alpha\beta}\},\{\sigma_{\alpha\beta\gamma}\})$ that satisfies following conditions \cite{GT10}: $\{U_{\alpha}\}$ is an open covering of a base space $X$, $L_{\alpha\beta}$ is a complex vector bundle over $U_{\alpha\beta}$,
and  $\sigma_{\alpha\beta\gamma}:L_{\alpha\beta}\otimes L_{\beta\gamma}\to L_{\alpha\gamma}$ is an isomorphism between complex vector bundles. They satisfy a commutative diagram
\begin{eqnarray}\label{eq:consistency}
  \begin{CD}
     L_{\alpha\beta}\otimes L_{\beta\gamma}\otimes L_{\gamma\delta} @>{1\otimes\sigma_{\beta\gamma\delta}}>> L_{\alpha\beta}\otimes L_{\beta\delta} \\
  @V{\sigma_{\alpha\beta\gamma}}VV    @V{\sigma_{\alpha\beta\delta}}VV \\
     L_{\alpha\gamma}\otimes L_{\gamma\delta}   @>{\sigma_{\alpha\gamma\delta}}>>  L_{\alpha\delta}.
  \end{CD}
\end{eqnarray}
It is known that gerbes on a topological space $X$ are classified by $\cohoZ{3}{X}$
\cite{Brylinski}.
This is 
a primary reason that we expect
a gerbe is an underlying mathematical
structure for parameterized $(1+1)$-dimensional invertible states and MPSs over $X$,
and,
by constructing a gerbe from a family of $(1+1)$-dimensional systems, we can extract a topological invariant that takes its value in $\cohoZ{3}{X}$.

\subsection{Definition of A Constant Rank MPS Gerbe}
\label{Definition of A Constant Rank MPS Gerbe}

Let's construct a gerbe on $X$ from a family of injective MPS matrices
parametrized by $X$.
For simplicity, we will first keep the rank (bond dimension) of MPSs constant. 
We will drop this condition later in Sec.\ \ref{sec:non-const_gerbe}.

To set the stage, 
we fix an open covering $\{U_{\alpha}\}_{\alpha\in I}$ of $X$
and consider $n\times n$ injective MPS matrices $\{A^{i}_{\alpha}(x)\}$ on each $U_{\alpha}$. 
At the intersection of two patches, 
$U_{\alpha\beta}$,
we have two MPSs representing the same physical state defined at $x \in X$.
By the fundamental theorem for (bosonic) MPSs,
these two MPSs are related by a gauge transformation, 
\begin{eqnarray}
A^{i}_{\alpha}(x)=g_{\alpha\beta}(x)A^{i}_{\beta}(x)g_{\alpha\beta}^{\dagger}(x),
\end{eqnarray}
where 
$g_{\alpha\beta}$ 
is an element of the projective unitary group, 
$g_{\alpha\beta}\in\mathrm{PU}(n)$\footnote{For simplicity, we omit the phase redundancy of MPSs.}. 
We call $g_{\alpha\beta}$ a transition function.
Let's take a $\mathrm{U}(n)$-lift $\{\hat{g}_{\alpha\beta}\}$ of $\{g_{\alpha\beta}\}$. 
From this unitary matrices $\{\hat{g}_{\alpha\beta}\}$, we define a state over $U_{\alpha\beta}$ by
\begin{eqnarray}
\ket{\{\hat{g}_{\alpha\beta}\}}
:=\sum_{\{i_{k}\}}\cdots A^{i_1}_{\alpha}\cdots A^{i_{p}}_{\alpha}\hat{g}_{\alpha\beta}A^{i_{p+1}}_{\beta}\cdots A^{i_{L}}_{\beta}\cdots\ket{\cdots i_{1}\cdots i_{i_{L}}\cdots}.
\end{eqnarray}
Here, because of a translation symmetry, the right-hand side does not depend on $p\in\Z$. 
Although this vector contains ambiguous infinite products 
in its coefficients, when calculating physical quantities 
(such as the higher Berry phase), as we will see below, 
we extract them by contracting the ends using suitable eigenvectors of suitable transfer matrices.
The state $\ket{\{\hat{g}_{\alpha\beta}\}}$
is reminiscent of the so-called mixed gauge MPS.
We also define a complex line bundle over $U_{\alpha\beta}$ by 
 \begin{eqnarray}
 L_{\hat{g}_{\alpha\beta}}:=\mathbb{C}\ket{\{\hat{g}_{\alpha\beta}\}}.
 \end{eqnarray}
Finally, on
a triple intersection $U_{\alpha\beta\gamma}$, we define an isomorphism
\begin{eqnarray}
 \sigma_{\alpha\beta\gamma}^{\rm MPS}:L_{\hat{g}_{\alpha\beta}}\otimes L_{\hat{g}_{\beta\gamma}}\to L_{\hat{g}_{\alpha\gamma}}:\ket{\{\hat{g}_{\alpha\beta}\}}\otimes\ket{\{\hat{g}_{\beta\gamma}\}}\mapsto\ket{\{\hat{g}_{\alpha\beta}\hat{g}_{\beta\gamma}\}}.
\end{eqnarray}
Let's check the commutative diagram Eq.\ (\ref{eq:consistency}) for $(\{U_{\alpha}\},\{L_{\hat{g}_{\alpha\beta}}\},\{\sigma_{\alpha\beta\gamma}^{\rm MPS}\})$: 
There exists $c_{\alpha\beta\gamma}\in\mathrm{U}(1)$ on $U_{\alpha\beta\gamma}$ 
so that
\begin{eqnarray}
    \hat{g}_{\alpha\beta}\hat{g}_{\beta\gamma}=c_{\alpha\beta\gamma}\hat{g}_{\alpha\gamma}.
\end{eqnarray}
Since 
$\ket{\{\hat{g}_{\alpha\beta}
\hat{g}_{\beta\gamma}\}}=c_{\alpha\beta\gamma}\ket{\{\hat{g}_{\alpha\gamma}\}}$, 
Eq.\ \eqref{eq:consistency} is equivalent to $(\delta c)_{\alpha\beta\gamma\delta}:=c_{\alpha\beta\gamma}c_{\alpha\beta\delta}^{\ast}c_{\alpha\gamma\delta}c_{\beta\gamma\delta}^{\ast}=1$\footnote{Here, $\delta$ is the coboundary operator of the \v{C}eck cohomology.},
and this equation follows from the associativity of the matrix product. 
Therefore, $(\{U_{\alpha}\},\{L_{\hat{g}_{\alpha\beta}}\},\{\sigma_{\alpha\beta\gamma}^{\rm MPS}\})$ 
is a gerbe on $X$. Since $(\delta c)_{\alpha\beta\gamma\delta}=1$, $c_{\alpha\beta\gamma}$ defines a $2$nd \v{C}eck cohomology class $\left[c_{\alpha\beta\gamma}\right]\in\coho{2}{X}{\underline{\mathrm{U}(1)}}\simeq\cohoZ{3}{X}$\footnote{Here, the isomorphism is given in the following way: Let's take a $\mathbb{R}$-lift $w_{\alpha\beta\gamma}$ of $c_{\alpha\beta\gamma}$, i.e., $c_{\alpha\beta\gamma}=e^{i2\pi w_{\alpha\beta\gamma}}$. Then, on $U_{\alpha\beta\gamma\delta}$,  $d_{\alpha\beta\gamma\delta}:=w_{\alpha\beta\gamma}-w_{\alpha\beta\delta}+w_{\alpha\gamma\delta}-w_{\beta\gamma\delta}$ takes its value in $\mathbb{Z}$ and satisfies the cocycle condition. Thus it defines the $3$rd cohomology class $\left[d_{\alpha\beta\gamma\delta}\right]\in\cohoZ{3}{X}$, and this is a topological invariant of the gerbe.}. $\left[c_{\alpha\beta\gamma}\right]$, which is a topological invariant of a gerbe, 
and
called the Dixmier-Douady class
\cite{DD63}. 
In the following, we call $(\{U_{\alpha}\},\{L_{\hat{g}_{\alpha\beta}}\},\{\sigma_{\alpha\beta\gamma}^{\rm MPS}\})$ a constant-rank MPS gerbe. Here, the adjective constant-rank implies the bond dimension of MPS matrices is constant over the parameter space $X$. 

A constant-rank MPS gerbe is a proper mathematical structure
to describe invertible states over $X$
when we are interested in 
a torsion part of
$\cohoZ{3}{X}$, i.e., 
a finite order subgroup of $\cohoZ{3}{X}$.
Such cases have been studied in detail in Ref.\ \cite{OTS23}.
In general, however, the rank of MPS matrices 
may not be constant
over the parameter space $X$
\cite{OSS22}.
Moreover, constant-rank MPS matrices cannot describe
nontrivial models which 
take their values in the free part,
i.e., 
(copies of) the infinite cyclic group 
$\mathbb{Z}$,
of $\cohoZ{3}{X}$.
Let us briefly explain this point. 
Since $\hat{g}_{\alpha\beta}\hat{g}_{\beta\gamma}=c_{\alpha\beta\gamma}\hat{g}_{\alpha\gamma}$ holds as a unitary matrix, 
the following equation is obtained by taking the determinant of both sides:
\begin{eqnarray}
    \det(\hat{g}_{\alpha\beta})\det(\hat{g}_{\alpha\gamma})^{\ast}\det(\hat{g}_{\beta\gamma})=c_{\alpha\beta\gamma}^{n}.
\end{eqnarray}
This equation implies that $c_{\alpha\beta\gamma}^{n}$ is closed cocycle and  $[c_{\alpha\beta\gamma}^{n}]$ is trivial in $\coho{2}{X}{\underline{\mathrm{U}(1)}}$, i.e., $[c_{\alpha\beta\gamma}^{n}]=1\in\coho{2}{X}{\underline{\mathrm{U}(1)}}$. Therefore, the topological class of $(\{U_{\alpha}\},\{L_{\hat{g}_{\alpha\beta}}\},\{\sigma_{\alpha\beta\gamma}^{\rm MPS}\})$ is in the torsion part of $\cohoZ{3}{X}$\footnote{We can also show this point using differential forms. By taking the logarithm, determinant, and exterior derivative of both sides of $\hat{g}_{\alpha\beta}\hat{g}_{\beta\gamma}=c_{\alpha\beta\gamma}\hat{g}_{\alpha\gamma}$,
\begin{eqnarray}\label{eq:gerbe_consistency}
d\log\det(\hat{g}_{\alpha\beta})-d\log\det(\hat{g}_{\alpha\gamma})+d\log\det(\hat{g}_{\beta\gamma})=d\log(c_{\alpha\beta\gamma}).
\end{eqnarray}
This implies $(w_\alpha:=0,d\log\det(\hat{g}_{\alpha\beta}),c_{\alpha\beta\gamma})$ is a $3$rd smooth Deligne cocycle
\cite{Brylinski,OTS23}. 
Since this cocycle is flat, i.e., $\eta_{\alpha}:=dw_\alpha=0$, the topological class of $(\{U_{\alpha}\},\{L_{\hat{g}_{\alpha\beta}}\},\{\sigma_{\alpha\beta\gamma}^{\rm MPS}\})$ is trivial in the free part of $\cohoZ{3}{X}$. This property is completely determined by the Dixmier-Douady class and independent of the choice of the higher connections.}.  This is due to the mathematical fact that the topological class of a $\mathrm{PU}(n)$-bundle can only take its value in the torsion part of $\cohoZ{3}{X}$
\cite{DK70}. 
Therefore, we need to handle a family of MPS matrices with a non-constant rank and construct a gerbe 
from such matrices\footnote{According to mathematics, 
another way to avoid this obstacle is to consider the case of $n=\infty$
\cite{AS05}. 
However, it is practically difficult to deal with MPSs of infinite rank.}. 
We discuss this point in Sec.\ \ref{sec:DD class form boomerang}.

\subsection{Triple Inner Product of MPSs}
\label{sec:DD class form boomerang}

Before delving into non-constant-rank MPSs, 
let us discuss one more ingredient, still using constant-rank MPSs. 
Specifically, we will demonstrate how the data that makes up the MPS gerbe, such as the transition functions and the Dixmier-Douady class, relate to certain overlaps of MPSs. We will show that the Dixmier-Douady class can be obtained from the triple inner product, defined below, for three MPSs. This is reminiscent of Wu-Yang's work on U(1) magnetic monopoles, where a topological invariant, the Chern class, can be obtained from the inner product of two wave functions from different patches. In this discussion, we present an alternative formulation in which the MPS gerbe's data is expressed in terms of (triple) wave function overlaps.
Moreover, in the following section, we'll see that this formulation also naturally generalizes to a definition of a gerbe from MPSs over $X$ with a non-constant rank.

\begin{figure}[t]
 \begin{center}
  \includegraphics[scale=1.5]{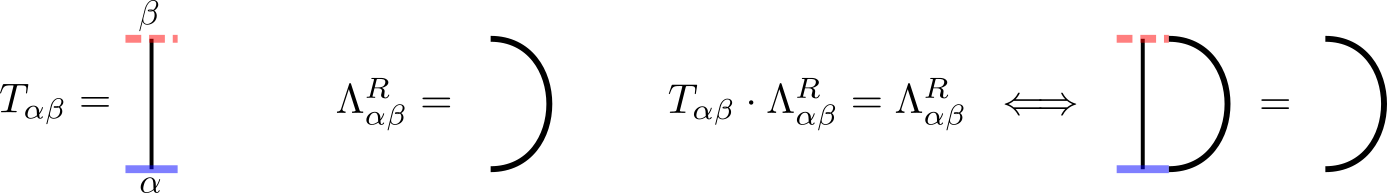}
 \end{center}
  \caption[]{
  The mixed transfer matrix $T_{\alpha\beta}$ and its right eigenvalue equation.
  }
\label{fig:eigenvalue eq}
\end{figure}

Let us start with 
the transfer matrix at $x\in U_\alpha$ is defined by 
\begin{eqnarray}
T_{\alpha}(x)=\sum_{i}A^{i\ast}_{\alpha}(x)\otimes A^{i}_{\alpha}(x).
\end{eqnarray}
As reviewed in Sec.\ \ref{Brief Review of MPS},
$T_{\alpha}(x)$ acts on $\mathrm{Mat}_{n}(\mathbb{C})$ from
the left and right as
$
T_{\alpha}(x)\cdot M:=\sum_{i}A^{i}_{\alpha}(x)M A^{i\dagger}_{\alpha}(x)
$,
$
M\cdot T_{\alpha}(x):=\sum_{i}A^{i\dagger}_{\alpha}(x)M A^{i}_{\alpha}(x)
$,
respectively, 
for arbitrary $M\in\mathrm{Mat}_{n}(\mathbb{C})$.
We represent this action pictorially 
as in Fig.\ \ref{fig: action of transfer matrix}.
The transfer matrix $T_{\alpha}(x)$ has unique right and left eigenvectors $\Lambda^{R}_{\alpha}(x)$ and $\Lambda^{L}_{\alpha}(x)$ with eigenvalue $1$:
\begin{align}
  T_{\alpha}(x)\cdot\Lambda^{R}_{\alpha}(x)=\Lambda^{R}_{\alpha}(x), 
  \quad
\Lambda^{L}_{\alpha}(x)\cdot T_{\alpha}(x)=\Lambda^{L}_{\alpha}(x).
\end{align}
A primary tool in this section is a mixed transfer matrix,
which we define from $\{A^i_{\alpha}(x)\}$
and $\{A^i_{\beta}(x)\}$ as
\begin{eqnarray}
T_{\alpha\beta}(x):=\sum_{i}A^{i\ast}_{\beta}(x)\otimes A^{i}_{\alpha}(x),
\end{eqnarray}
over $U_{\alpha\beta}$. A crucial point is that the spectrum 
of $T_{\alpha\beta}(x)$ is identical to 
that of $T_{\alpha}(x)$, and in particular, $T_{\alpha\beta}(x)$ has
unique left and right eigenvectors with eigenvalue $1$.
Let's check this point. 
From now on, we omit the dependence on $x$.
Let $\Lambda_{\alpha}^{R,k}$ be the $k$-th eigenvector of $T_{\alpha}$ with eigenvalue $\lambda_{\alpha}^{R,k}$,
$
T_{\alpha}\cdot\Lambda_{\alpha}^{R,k}=\lambda_{\alpha}^{R,k}\Lambda_{\alpha}^{R,k}.
$
Then $\Lambda_{\alpha\beta}^{R,k}:=\Lambda_{\alpha}^{R,k}\hat{g}_{\alpha\beta}$ is the eigenvector of $T_{\alpha\beta}$ with the same eigenvalue $\lambda_{\alpha}^{R,k}$:
\begin{align}
  T_{\alpha\beta}\cdot\Lambda_{\alpha\beta}^{R,k}
  =\sum_{i}A^{i}_{\alpha}(\Lambda_{\alpha}^{R,k}\hat{g}_{\alpha\beta})(\hat{g}_{\beta\alpha}A^{i\dagger}_{\alpha}\hat{g}_{\beta\alpha}^{\dagger})
  =\lambda_{\alpha}^{R,k}\Lambda_{\alpha}^{R,k}\hat{g}_{\beta\alpha}^{\dagger}
  =\lambda_{\alpha}^{R,k}\Lambda_{\alpha\beta}^{R,k}.
\end{align}
Therefore, there is a one-to-one correspondence between the eigenvectors of $T_{\alpha}$ and $T_{\alpha\beta}$ with the same eigenvalue. Similarly, for a left eigenvector $\Lambda_{\alpha}^{L,k}$ of $T_{\alpha}$ with eigenvalue $\lambda_{\alpha}^{L,k}$, $\Lambda_{\alpha\beta}^{L,k}:=\hat{g}_{\beta\alpha}\Lambda_{\alpha}^{L,k}$ is a left eigenvector of $T_{\alpha\beta}$ with the eigenvalue $\lambda_{\alpha}^{L,k}$:
\begin{align}
  \Lambda_{\alpha\beta}^{L,k}\cdot T_{\alpha\beta}
  &=\sum_{i}(\hat{g}_{\beta\alpha}A^{i\dagger}_{\alpha}\hat{g}_{\beta\alpha}^{\dagger})\hat{g}_{\beta\alpha}\Lambda_{\alpha}^{L,k} A^{i}_{\alpha}
  =\lambda_{\alpha}^{L,k}\hat{g}_{\beta\alpha}\Lambda_{\alpha}^{L,k}
=\lambda_{\alpha}^{L,k}\Lambda_{\alpha\beta}^{L,k}.
\end{align}
We define the right and left eigenstates
of $T_{\alpha\beta}$ with eigenvalue $1$ by
\begin{eqnarray}
  \Lambda_{\alpha\beta}^{R}:=\Lambda_{\alpha}^{R}\hat{g}_{\alpha\beta}=\hat{g}_{\alpha\beta}\Lambda_{\beta}^{R},
                                \quad
\Lambda_{\alpha\beta}^{L}:=\hat{g}_{\beta\alpha}\Lambda_{\alpha}^{L}=\Lambda_{\beta}^{L}\hat{g}_{\beta\alpha}.
\end{eqnarray}
We represent the eigenvalue equations pictorially as in Fig.\ \ref{fig:eigenvalue eq}.
In the right canonical form, 
$\Lambda^{R}_{\alpha}=1_{n}$. 
We also fix the phase of $\Lambda^{L}_{\alpha}$ by the condition $\tr{\Lambda^{L}_{\alpha}}=1$. 
This is the normalization condition of the infinite MPS. 
Remark that the phases of $\Lambda_{\alpha\beta}^{R}$ and $\Lambda_{\alpha\beta}^{L}$ are still redundant, 
but the redefinition of them can be absorbed in the $\mathrm{U}(n)$-lift of the transition functions. 

\begin{figure}[t]
 \begin{center}
  \includegraphics[scale=1.2]{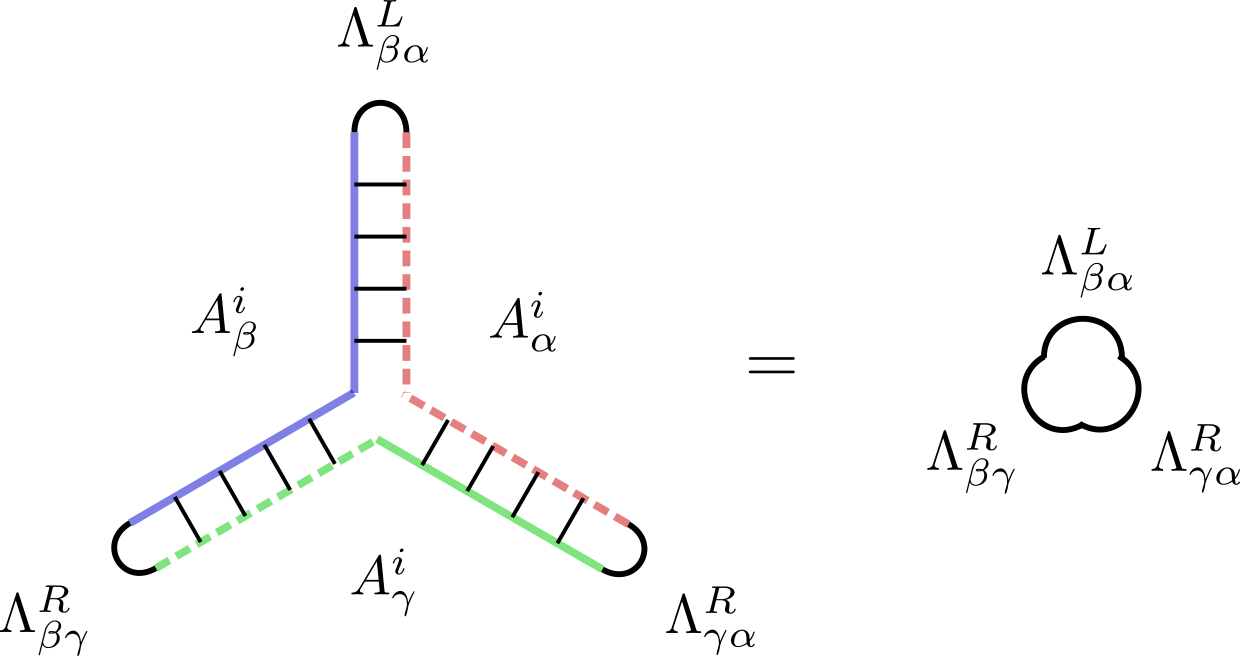}
 \end{center}
  \caption[]{
  The triple inner product of 
  three MPSs 
  $\{A^i_{\alpha}\}$,
  $\{A^i_{\beta}\}$,
  $\{A^i_{\gamma}\}$
  from different patches 
  $U_{\alpha}, U_{\beta}, U_{\gamma}$, 
  respectively.
  The right side of the middle wing and the right side of the bottom right wing represent 
  the matrix $A^{i}_{\alpha}$, 
  the left side of the middle wing and the left side of the bottom left wing 
  represent the matrix $A^{i}_{\beta}$, 
  and 
  the right side of the bottom left wing and the left side of the bottom right wing represent 
  the matrix $A^{i}_{\gamma}$. 
  The dotted lines represent the complex conjugation of the MPS matrices.
}
\label{fig:boomerang}
\end{figure}

We are now ready to define the triple inner product. 
On a triple intersection $U_{\alpha\beta\gamma}$, 
consider the "boomerang" diagram as in Fig.\ \ref{fig:boomerang}. 
Here, three infinite MPSs,
representing the same physical state at $x\in X$,
from three different patches are "glued" 
together as in Fig.\ \ref{fig:boomerang}. 
Observe how "bra" and "ket" MPS matrices are
arranged depending on which "wing" they are located.
At the infinities of the three "wings", the tensor network is capped off by 
putting either left or right eigenvectors.
The products of the mixed transfer matrices 
are easily computed in the thermodynamic limit, and 
we can check
that the boomerang diagram computes 
the Dixmier-Douady class:
\begin{eqnarray}
\text{the boomerang diagram}=\tr{\Lambda_{\beta\alpha}^{L}\Lambda_{\beta\gamma}^{R}\Lambda_{\gamma\alpha}^{R}}=\tr{\Lambda_{\alpha}^{L}\hat{g}_{\alpha\beta}1_{n}\hat{g}_{\beta\gamma}1_{n}\hat{g}_{\gamma\alpha}}
=c_{\alpha\beta\gamma}.
\end{eqnarray}
We define the triple inner product of three MPSs as the "boomerang" diagram and the higher Berry phase as the Dixmier Douady class. The ordinary Berry phase can be obtained from the ordinary inner product of two wavefunctions that are physically 
the same but taken from two different patches. As the natural generalization of this method, the higher Berry phase in $(1+1)$-dimensional systems can be obtained from the triple inner product of three MPSs that are physically the same but taken from three different patches.
Note that with the mixed transfer matrix and 
the triple inner product, 
it is not necessary to deal with  
the transition functions explicitly.
Instead, the data necessary to define the (constant-rank) MPS gerbe are encoded in the mixed transfer matrix and 
the triple inner product.

Finally, we
note that there are some ambiguities in the definition of an MPS gerbe and a triple inner product. For example, a gerbe can be constructed by using $\Lambda^{L}_{\beta\alpha}$ instead of $\Lambda^{R}_{\alpha\beta}$ in the definition of the line bundle on $U_{\alpha\beta}$.
In our choice,
$|\{A^i_{\alpha}\}\rangle$,
$|\{\hat{g}_{\alpha\beta} \}\rangle$
and 
the modulus of the triple inner product 
are all normalized to be one,
while in other choices we would
need to adjust   
normalization 
(by properly rescaling 
$\Lambda^L_{\beta\alpha}$). 
Our choice would be natural in this sense.
%\delete{, but in this case we need to adjust the normalization of $\Lambda^{L}_{\beta\alpha}$. Our choice is natural in the sense that the infinite MPS is normalized.} 

\subsection{Definition of A Non-Constant Rank MPS Gerbe}\label{sec:non-const_gerbe}

In Secs.\  \ref{Definition of A Constant Rank MPS Gerbe}
and \ref{sec:DD class form boomerang},
we assume that the rank of the MPS matrices is constant over the parameter space $X$. As a generalization of this situation, we consider a family of MPS matrices with non-constant rank. To this end, we first introduce a notion of essentially injective matrices: 
let $\{A^i\}$ be 
a set of $n\times n$ matrices. Then $\{A^i\}$ is essentially injective if and only if there is an invertible matrix $X$ such that 
\begin{eqnarray}\label{eq:ess_right_canonical}
XA^{i}X^{-1}=\begin{pmatrix}
\tilde{A}^{i}& 0\\
Y^i&0
\end{pmatrix},
\end{eqnarray}
for some $\tilde{n}\times\tilde{n}$ injective matrices $\{\tilde{A}^{i}\}$ and $(n-\tilde{n})\times \tilde{n}$ matrices $Y^{i}$.
Also, we impose the right canonical form condition,
$\sum_i A^{i} A^{i\dagger} = 1_n$. 
In terms of $\{\tilde{A}^i\}$ and $Y_i$, 
this means that
\begin{align}
\label{right cano cond}
&
\sum_i \tilde{A}^i \tilde{A}^{i\dagger} = 1_{\tilde{n}},
\quad 
\sum Y^i Y^{i\dagger}= 1_{n-\tilde{n}},
\nonumber \\
&
\sum_i Y^i \tilde{A}^{i\dagger} =0,
\quad
\sum_i \tilde{A}^{i} Y^{i\dagger}=0.
\end{align}
We call $\tilde{n}$ an essential rank of the essentially injective matrices,
and $\{\tilde{A}^i\}$ the injective part of the essentially injective matrices. 
Usually, we eliminate the lower triangular component $Y^i$ by hand because it does not affect the state. However, such cases appear naturally when considering a family of MPS matrices.

Let $\{U_{\alpha}\}$ be an open covering of $X$ and let's consider a family of essentially injective MPS matrices. Assume that the rank of MPS matrices is constant on each patch. Let $\{A^{i}_{\alpha}\}$ be  $n_{\alpha}\times n_{\alpha}$ essentially injective matrices whose essential rank  $\tilde{n}_{\alpha}(x)$ 
can be dependent on $x\in U_{\alpha}$. 
We also assume that $\tilde{n}_{\alpha}(x)=\tilde{n}_{\beta}(x)$ on non-empty intersection $U_{\alpha\beta}$.
Let's consider the mixed transfer matrix
\begin{eqnarray}
T_{\alpha\beta}=\sum_{i}A^{i\ast}_{\beta}\otimes A^{i}_{\alpha}.
\end{eqnarray} 
The mixed transfer matrix $T_{\alpha\beta}$ acts on $M\in \mathrm{Mat}_{n_{\alpha}\times n_{\beta}}(\mathbb{C})$ 
from the left 
as
\begin{align}
 T_{\alpha\beta}\cdot M= \sum_{i}A^{i}_{\alpha}M A^{i\dagger}_{\beta},
\end{align}
and acts on $M\in \mathrm{Mat}_{n_{\beta}\times n_{\alpha}}(\mathbb{C})$ from the right as
\begin{align}
 M \cdot T_{\alpha\beta}= \sum_{i}A^{i\dagger}_{\beta}M A^{i}_{\alpha}.
\end{align}
Then we can show that both the maximal left and right eigenvalues of the mixed transfer matrix are $1$, 
and the right and left eigenvectors $\Lambda_{\alpha\beta}^{R}$ and $\Lambda_{\alpha\beta}^{L}$ are unique and 
given by
\begin{eqnarray}
\label{eq: right eig vec}
    \Lambda_{\alpha\beta}^{R}:=\begin{pmatrix}
        \tilde{\Lambda}_{\alpha\beta}^{R}&  0
        \\
        0 &
        \sum_i Y^i_{\alpha} 
         \tilde{\Lambda}^R_{\alpha\beta} Y^{i\dagger}_{\beta} 
    \end{pmatrix}
    \quad
\mbox{and}
\quad
    \Lambda_{\alpha\beta}^{L}:=\begin{pmatrix}
        \tilde{\Lambda}_{\alpha\beta}^{L}& 0 \\
        0 & 0 
    \end{pmatrix},
\end{eqnarray}
respectively, 
where $\tilde{\Lambda}_{\alpha\beta}^{R}$ and $\tilde{\Lambda}_{\alpha\beta}^{L}$ are right and left eigenvectors with eigenvalue $1$ of the mixed transfer matrix of injective part of $\{A^i_{\alpha}\}$ and $\{A^{i}_{\beta}\}$.

This can be readily checked 
as follows.
Let $M$ be an $n_{\alpha}\times n_{\beta}$ matrix and consider the following decomposition:
\begin{eqnarray}
\label{decomp}
M=\begin{pmatrix}
    \Lambda&Z\\
    X&\Lambda'
\end{pmatrix},
\end{eqnarray}
where 
$\Lambda$, $X$, $Z$, and $\Lambda'$ 
are $\tilde{n}_{\alpha}\times \tilde{n}_{\alpha}$, 
$(n_{\alpha}-\tilde{n}_{\alpha})\times \tilde{n}_{\alpha}$,
$\tilde{n}_{\alpha}\times (n_{\beta}-\tilde{n}_{\alpha})$,
and $(n_{\alpha}-\tilde{n}_{\alpha})\times(n_{\beta}-\tilde{n}_{\alpha})$, respecitvely.
Then, the right eigenvalue equation 
$T_{\alpha\beta}\cdot M=M$ reads
\begin{align}
\sum_i
\begin{pmatrix}
\tilde{A}^i_{\alpha}\Lambda \tilde{A}^{i\dagger}_{\beta} &
\tilde{A}^i_{\alpha}\Lambda Y^{i\dagger}_{\beta} 
\\
Y^{i}_{\alpha}\Lambda \tilde{A}^{i\dagger}_{\beta}
&
Y^i_{\alpha} \Lambda Y^{i\dagger}_{\beta}
\end{pmatrix}
=
\begin{pmatrix}
    \Lambda & Z \\ X & \Lambda'
\end{pmatrix}.
\end{align}
From the upper left block, we see that $\Lambda$ must be the right eigenvector, 
$\Lambda=\Lambda^R_{\alpha\beta}=g_{\alpha\beta}$.
We also see from the lower left block
$
\sum_i Y^i_{\alpha} \Lambda 
\tilde{A}^{i\dagger}_{\beta}
=
\sum_i Y^i_{\alpha} g_{\alpha\beta}  
\tilde{A}^{i\dagger}_{\beta}
=
\sum_i Y^i_{\alpha}  
\tilde{A}^{i\dagger}_{\alpha}
g^{\dagger}_{\alpha\beta}
=
0
$
where we used the right canonical condition
\eqref{right cano cond}.
We can show similarly 
that 
$\sum_i \tilde{A}^{i}_{\alpha}\Lambda Y^{i\dagger}_{\beta}
=
g_{\alpha\beta}
\sum_i \tilde{A}^i_{\beta} Y^{i\dagger}_{\beta}
=0
$.
We thus conclude the first equation in \eqref{eq: right eig vec}.
For the left eigenequation
$
M\cdot T_{\alpha\beta}=M
$, 
we consider the similar decomposition
\eqref{decomp},
which leads to 
\begin{align}
\sum_i
\begin{pmatrix}
(\tilde{A}^{i\dagger}_{\beta} \Lambda +
Y^{i\dagger}_{\beta} X) \tilde{A}^i_{\alpha}
+
(\tilde{A}^{i\dagger}_{\beta}
Z + Y^{i\dagger}_{\beta}\Lambda') Y^{i}_{\alpha}
& 0 \\
0 & 0
   \end{pmatrix}
   =
 \begin{pmatrix}
 \Lambda & Z \\
 X &\Lambda'
   \end{pmatrix}.
\end{align}
The solution is given by
the second equation in \eqref{eq: right eig vec},
$X=Z=\Lambda'=0$
and 
$\Lambda=\tilde{\Lambda}^L_{\alpha\beta}$.

We are now ready to define 
a gerbe from a family of essentially injective matrices, including the case where the rank 
is not constant over the parameter space. 
It is defined, as a natural generalization of a constant-rank MPS gerbe,
as follows:
We define a state over $U_{\alpha\beta}$ by
\begin{eqnarray}
\ket{\{\Lambda^{R}_{\alpha\beta}\}}:=\sum_{\{i_{k}\}}\cdots A^{i_1}_{\alpha}\cdots A^{i_{p}}_{\alpha}\Lambda^{R}_{\alpha\beta}A^{i_{p+1}}_{\beta}\cdots A^{i_{L}}_{\beta}\cdots\ket{\cdots i_{1}\cdots i_{i_{L}}\cdots},
\end{eqnarray}
and a complex line bundle over $U_{\alpha\beta}$ by 
\begin{eqnarray}
    L^{\rm MPS}_{\alpha\beta}:=\mathbb{C}\ket{\{\Lambda^{R}_{\alpha\beta}\}}.
\end{eqnarray}
On a triple intersection, we define an isomorphism
\begin{eqnarray}
 \sigma_{\alpha\beta\gamma}^{\rm MPS}:L^{\rm MPS}_{\alpha\beta}\otimes L^{\rm MPS}_{\beta\gamma}\to L^{\rm MPS}_{\alpha\gamma}:\ket{\{\Lambda^{R}_{\alpha\beta}\}}\otimes\ket{\{\Lambda^{R}_{\beta\gamma}\}}\mapsto\ket{\{\Lambda^{R}_{\alpha\beta}\Lambda^{R}_{\beta\gamma}\}}=c_{\alpha\beta\gamma}\ket{\{\Lambda^{R}_{\alpha\gamma}\}}.
\end{eqnarray}
Then, 
$\mathcal{G}^{\rm MPS}:=(\{U_{\alpha}\},\{L^{\rm MPS}_{\alpha\beta}\},\{\sigma^{\rm MPS}_{\alpha\beta\gamma}\})$ 
is a gerbe on $X$. 
We call $\mathcal{G}^{\rm MPS}$ 
a non-constant-rank MPS gerbe or
an MPS gerbe for short. 
The triple inner product can also be defined
following the constant-rank case.  
We can compute the Dixmier-Douady class 
by the same diagram as in the Fig.\ \ref{fig:boomerang}:
\begin{eqnarray}\label{eq:general dd}
\text{the boomerang diagram}=\tr{\Lambda_{\beta\alpha}^{L}\Lambda_{\beta\gamma}^{R}\Lambda_{\gamma\alpha}^{R}}
=c_{\alpha\beta\gamma}.
\end{eqnarray}
Since $\{\Lambda^{L}_{\alpha\beta}\}$ 
includes the projection onto the injective part
and $\{\Lambda^{R}_{\alpha\beta}\}$ is block-diagonal, Eq.\ (\ref{eq:general dd}) 
reduces to
\begin{eqnarray}
\tr{\Lambda_{\beta\alpha}^{L}\Lambda_{\beta\gamma}^{R}\Lambda_{\gamma\alpha}^{R}}=\tr{\tilde{\Lambda}_{\beta\alpha}^{L}\tilde{\Lambda}_{\beta\gamma}^{R}\tilde{\Lambda}_{\gamma\alpha}^{R}}.
\end{eqnarray}
Namely, $c_{\alpha\beta\gamma}$ is nothing but the Dixmier-Douady class for the MPS matrices projected onto the injective part.

\section{Star Product and Integration}

\begin{figure}[t]
  \centering
  \includegraphics[scale=1.2]{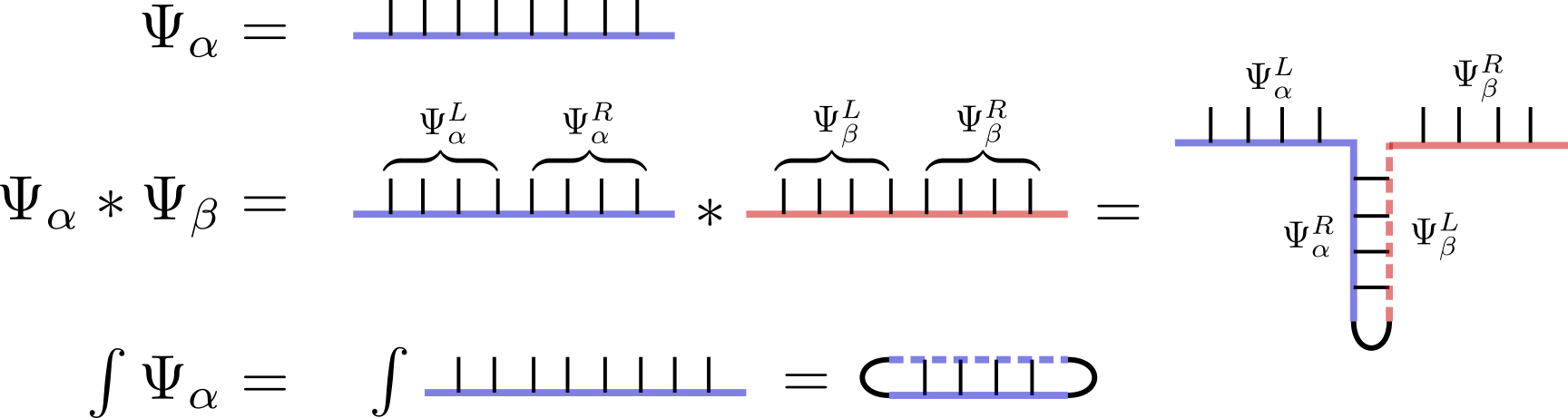}
  \caption{Matrix product states,
    $*$ product and integration.
    Along the dotted lines, the relevant MPS tensors are conjugated, i.e., $A^*$.}
  \label{fig:MPS1}
\end{figure}

In this section, we introduce two operations
for infinite MPSs -- 
star product ($*$) and integration ($\int$). 
As we will see, 
these operators are useful for describing 
the structures introduced in the preceding sections.
Our definitions 
are largely inspired by,
and 
essentially identical to,
the non-commutative geometry 
in string field theory
\cite{Witten:1985cc}.

Let us first introduce a multiplication law $*$ 
for two infinite MPSs
(Fig.\ \ref{fig:MPS1}).
In this section, 
we denote an MPS 
constructed from 
$\{A^i_{\alpha}\}$
as $\Psi_{{\alpha}}$.
For two MPSs $\Psi_{{\alpha}}$ 
and $\Psi_{{\beta}}$
from different patches
$U_{\alpha}$ and $U_{\beta}$, 
the product 
$\Psi_{{\alpha}} *  \Psi_{{\beta}}$ 
is defined by first splitting 
$\Psi_{{\alpha}}$ and 
$\Psi_{{\beta}}$ into their left and right pieces, denoted by
$\Psi^L_{\alpha}, \Psi^R_{\alpha}$ 
and $\Psi^L_{\beta}$ and 
$\Psi^R_{\beta}$, respectively.
In the product 
$\Psi_{{\alpha}} * \Psi_{{\beta}}$, 
$\Psi^R_{\alpha}$ and $\Psi^L_{\beta}$ are "glued",
i.e., contracted.
In this process, the MPS matrices 
$\{A^i_{\beta}\}$ on the left part of $\Psi_{\beta}$ are first converted 
to 
their conjugates 
$\{A^{i*}_{\beta}\}$ ("bras")
and then contracted with the right part of $\Psi_{\alpha}$.
The star product is associative,
$
(\Psi_{{\alpha}}* \Psi_{{\beta}}) * 
\Psi_{{\gamma}}
=
\Psi_{{\alpha}}* (\Psi_{{\beta}} *  \Psi_{{\gamma}})
$,
but not commutative. 
Intuitively,
we regard 
physical indices in 
$\Psi^L_{\alpha}$ 
and $\Psi^R_{\alpha}$ as
row (input) and column (output) 
indices of an infinite matrix,
or 
a semi-infinite matrix product operator.
Accordingly, 
the $*$ product 
can be interpreted as 
matrix multiplication
of two infinite-dimensional matrices.

To see the connection with the MPS gerbe,
we consider three MPSs 
$\Psi_{{\alpha}}, \Psi_{{\beta}}, \Psi_{{\gamma}}$ defined on patches
$U_{\alpha}, U_{\beta}, U_{\gamma}$,
respectively.
First, we can readily check that the product
$\Psi_{{\alpha}}* \Psi_{{\beta}}$ 
is nothing but
the mixed gauge MPS, 
$|\{\Lambda^R_{\alpha\beta} \}\rangle$.
Following the notation of this section,
we simply write 
$|\{\Lambda^R_{\alpha\beta}\}\rangle \equiv \Psi_{{\alpha\beta}}$.
We also note 
that 
an infinite canonical MPS is an idempotent of the $*$ product,
$\Psi_{{\alpha}} * \Psi_{{\alpha}}= \Psi_{{\alpha}}$.
Second,
the product of
$\Psi_{{\alpha\beta}}$ and $\Psi_{{\beta\gamma}}$ is given by
\begin{align}
\Psi_{{\alpha\beta}}* \Psi_{{\beta\gamma}}=
\Psi_{{\alpha}}
* 
\Psi_{{\beta}} * 
\Psi_{{\beta}}
*
\Psi_{{\gamma}}
=
c_{\alpha\beta\gamma}\Psi_{{\alpha\gamma}}.
\end{align}
Hence, the $*$ product is nothing but $\sigma^{{\rm MPS}}_{\alpha\beta\gamma}$.
We note that
mixed gauge MPSs are closed
under the multiplication $*$.

\begin{figure}[t]
  \centering
  \includegraphics[scale=1.2]{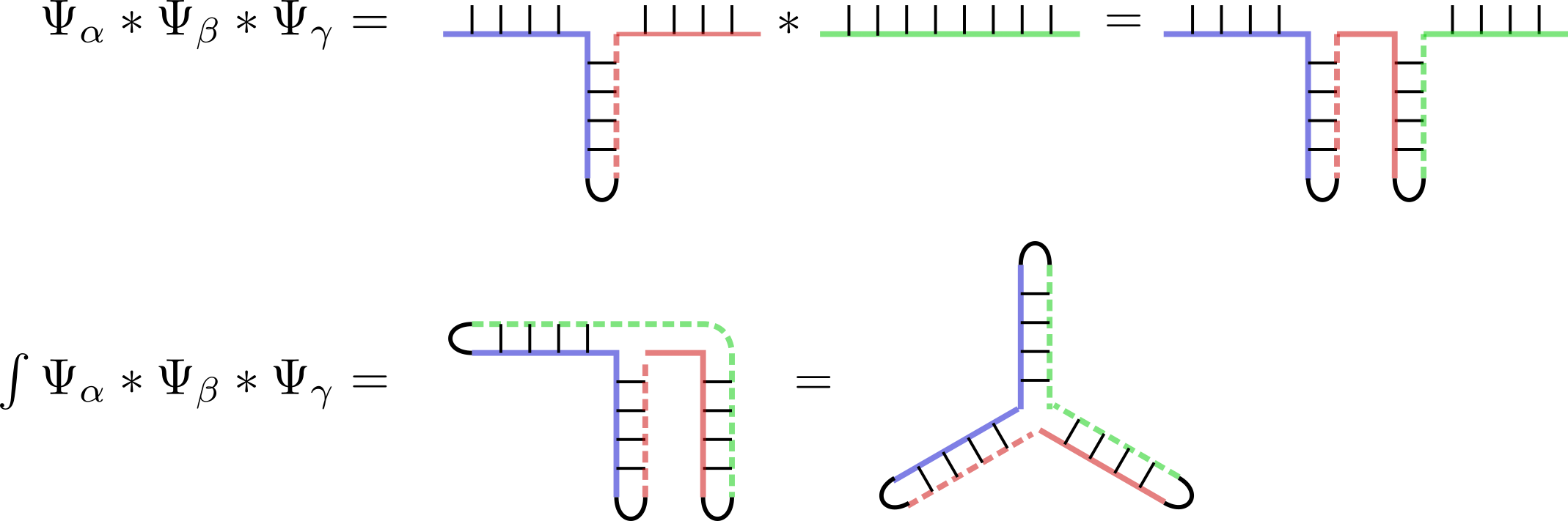}
  \caption{
  The star product of three MPSs
  $\Psi_{\alpha}$, $\Psi_{\beta}$, and $\Psi_{\gamma}$,
    and the triple inner product.}
  \label{fig:MPS2}
\end{figure}

To see how the triple inner product arises, 
we also introduce an "integration" $\int$.
To define the integration of $\Psi_{\alpha}$,
$\int \Psi_{\alpha}$, we "fold" 
$\Psi_{{\alpha}}$ and contract 
$\Psi^L_{\alpha}$ and $\Psi^R_{\alpha}$
(Fig.\ \ref{fig:MPS1}).
With this rule, we can see, for example, 
\begin{align}
    \int\,
    \displaystyle
    \Psi_{{\alpha}} &
    =
    \mathrm{tr}\,
    (
    \Lambda^L_{\alpha} \Lambda^R_{\alpha}
    )
    =
    1,
  \nonumber \\
    \int
    \Psi_{{\alpha}}* 
    \Psi_{{\beta}} 
    &
    =
    \mathrm{tr}\,
    (\Lambda^L_{\alpha\beta}
    \Lambda^R_{\alpha\beta})
    =
     \mathrm{tr}\,
    (\Lambda^L_{\beta} 
    \hat{g}_{\beta \alpha}
    \hat{g}_{\alpha \beta}\Lambda^R_{\beta})
    =
    1.
\end{align}
Namely, $\int \Psi_{{\alpha}}$ is 
the norm of $\Psi_{{\alpha}}$
and 
$\int \Psi_{{\alpha}} * \Psi_{{\beta}}$
is the overlap of 
$\Psi_{{\alpha}}$ and 
$\Psi_{{\beta}}$.
It is also evident 
that 
$
    \int \Psi_{{\alpha}} * \Psi_{{\beta}}
    =
    \int \Psi_{{\beta}} * \Psi_{{\alpha}}
$.
As before,
regarding
the physical indices in 
$\Psi^L_{\alpha}$ 
and $\Psi^R_{\alpha}$ as
row and column 
indices,
the integration 
is interpreted as 
the matrix trace. 
Finally, we can readily see that 
the integral of the triple product
$\Psi_{{\alpha}}* \Psi_{{\beta}} * \Psi_{{\gamma}}$
is the triple inner product,  
\begin{align}
    \int
    \Psi_{\alpha} * \Psi_{\beta} * \Psi_{\gamma}
    =
    c_{\alpha\beta\gamma}.
\end{align}
The cyclicity
$
\int \Psi_{\alpha} * \Psi_{\beta} * \Psi_{\gamma}
=
\int \Psi_{\beta} * \Psi_{\gamma} * \Psi_{\alpha}
$
is evident from the cyclicity of
$c_{\alpha\beta\gamma}$.
Thus, the $*$ product 
and integration
reproduce the essential ingredients
of the MPS gerbe.
We note that the triple inner product 
can also be viewed as 
the regular inner product 
of two non-uniform states, 
$\Psi_{\alpha\beta}$
and 
$\Psi_{\beta\gamma}$.

Before leaving this section,
several comments are in order.

--
It appears that there is some flexibility 
in the definition of the $*$ product and the integration. 
For example, when we glue two 
MPSs $\Psi_{\alpha}$ and $\Psi_{\beta}$,
we can take the conjugate of 
$\Psi^R_{\alpha}$
while keeping 
$\Psi^L_{\beta}$ intact. 
As for the integration, we also have at least two choices, 
i.e., 
taking the conjugation of
$\Psi^L_{\alpha}$ or $\Psi^R_{\alpha}$.
To be consistent with 
the "regular rule" of matrix multiplication and trace, 
one would choose to 
take the conjugate of $\Psi^R_{\alpha}$
both in $\Psi_{\alpha}* \Psi_{\beta}$
and $\int \Psi_{\alpha}$;
in this convention, 
the left (right) part of an MPS
is always regarded as 
row (column) indices
(both in the $*$ product and trace).
This choice results in the different 
definition of 
an MPS gerbe and a triple inner product
as noted 
at the end of Sec.\ 
\ref{sec:DD class form boomerang}.
(The idempotent property $\Psi_{\alpha}*\Psi_{\alpha}=\Psi_{\alpha}$ however is 
lost in this choice.)
We also note that, 
while we 
have focused on 
the right canonical form,
we can adopt a different 
canonical form,
the mixed canonical form,
in particular.

-- The notations and ideas behind these definitions
are from noncommutative geometry 
\cite{Connes:1994yd}
--
$\Psi_{\alpha}, \Psi_{\beta}, \cdots$ can be thought of as an analog of differential forms, 
and the $*$ product is an analog of 
the wedge product. 
As differential forms, we should be able to integrate $\Psi_{\alpha}$.
The $*$ product and integration
are parts of the ingredients that constitute non-commutative geometry.
To fully define a non-commutative geometry, we need additional  
structures, the derivative, and $\mathbb{Z}_2$ grading.
In string field theory, the derivative is given by
the so-called BRST operator that is used to select physical states.
The $\mathbb{Z}_2$ grading is provided by the number of ghosts.
While we do not need such structures for the purpose of this paper,
i.e., to discuss the topological properties of 
gapped translationally-invariant ground states,
we may speculate that
the full non-commutative geometry structure may be useful 
once we consider a wider class of states, e.g., excited states.

-- 
We noted that an infinite 
MPS is an idempotent of the $*$ product,
i.e., projector,
$\Psi_{{\alpha}}* \Psi_{{\alpha}}= \Psi_{{\alpha}}$.
This is similar to the fact that
in string field theory, 
the matter part of 
the full string field satisfies
the same equation
\cite{Rastelli_2001, Kosteleck__2001, rastelli2001classical, Gross_2001},
and describes 
a 
D-brane (D25-brane) --
an extended object in string theory.
This is reminiscent of the fact 
that
invertible states in $(1+1)$ dimensions 
can be expressed as boundary states 
in boundary conformal field theory
\cite{Cho_2017}.
Furthermore, a mixed gauge MPS $\Psi_{\alpha\beta}$ can be interpreted 
as a boundary condition changing operator
\cite{Cardy:1986gw, Cardy:1989ir},
and the $*$ product 
$\Psi_{\alpha\beta}* \Psi_{\beta\gamma}
=
c_{\alpha\beta\gamma}\Psi_{\alpha\gamma}$
represents
the fusion of two boundary condition 
changing operators. 
With a proper regularization 
(Euclidean evolution),
the triple inner product 
corresponds to the partition function 
on a strip with 
boundary conditions 
specified by 
$\alpha, \beta$ and $\gamma$,
i.e., with an insertion of a boundary condition 
changing operator between $\alpha$ and $\beta$, say\footnote{
To describe a parameterized 
family of invertible states 
we expect that
these boundary conditions 
preserve only  
the conformal symmetry but not any larger symmetry.}.

\section{Discussion}

In this paper, we identified a gerbe structure 
for a family of infinite MPSs over a parameter space $X$. 
We also introduced, 
as a generalization of the ordinary Berry phase
for overlaps of two wavefunctions,
the triple inner product 
for three infinite MPSs and showed that 
it extracts the Dixmir-Douady class,
which is  
a topological invariant of 
an MPS gerbe and hence 
a family of invertible states over $X$.
Our formalism works both for 
the torsion and free parts of 
$\cohoZ{3}{X}$. 
In particular, for the free case, 
we showed how to handle 
non-constant rank MPSs over $X$. 

The relation between the triple inner product and 
the Dixmir-Douady class is one of the upshots of the paper.
In principle, this relation 
can provide a practical way to calculate 
the topological invariant for a given family of 
$(1+1)$-dimensional invertible states.
It would be an important next step to find an explicit "algorithm" for this and study examples.

In addition,
it is interesting to consider
the triple inner product of 
a larger class of MPSs, 
such as  
finite, and/or 
non-translationally invariant MPSs.
In particular, 
it may be interesting to study finite MPSs with 
periodic boundary conditions. 
We also note that 
a wave function overlap
for three many-body states,
similar to our triple inner product,
has been discussed 
as a numerical tool to 
extract universal data of 
$(1+1)$-dimensional 
lattice quantum systems
at criticality
\cite{Zou_2022,Zou_2022b,liu2022operator}.

\section*{Acknowledgements}

We thank useful discussions with Kiyonori Gomi, 
Yichen Hu, Yuya Kusuki, Yuhan Liu, Yoshiko Ogata and Ken Shiozaki. 
We thank the Yukawa Institute for Theoretical Physics at Kyoto University, where this work was initiated during the YITP-T-22-02 on "Novel Quantum States in Condensed Matter 2022".
S.O. was supported by the establishment of university fellowships towards the creation of science technology innovation.
S.R.~is supported by the National Science Foundation under 
Award No.\ DMR-2001181, and by a Simons Investigator Grant from
the Simons Foundation (Award No.~566116).
This work is supported by
the Gordon and Betty Moore Foundation through Grant
GBMF8685 toward the Princeton theory program.

\bibliography{ref}

\end{document}